\documentclass[10pt,conference]{IEEEtran}

\usepackage{times,amsmath,epsfig}
\usepackage{latexsym,amssymb}
\usepackage{cite}

\newcommand{\no}{\nonumber}
\newtheorem{define}{Definition}

\newcommand{\beq}{\begin{equation}}
\newcommand{\enq}{\end{equation}}
\newcommand{\beqa}{\begin{eqnarray}}
\newcommand{\enqa}{\end{eqnarray}}

\newcommand{\beql}[1]{\begin{equation}\label{#1}}

\newtheorem{thm}{Theorem}

\IEEEoverridecommandlockouts
\begin{document}


\title{Secrecy Capacity of the Wiretap Channel with Noisy Feedback}




\author{\authorblockN{Lifeng Lai\authorrefmark{1},
Hesham El Gamal\authorrefmark{2}, H. Vincent
Poor\authorrefmark{3}\thanks{This research was supported by the
National Science Foundation under Grants ANI-03-38807 and
CNS-06-25637.}}
\authorblockA{\authorrefmark{1}\authorrefmark{3} Department of Electrical Engineering\\
Princeton University, Princeton, NJ 08512, USA\\
Email: \{llai,poor\}@princeton.edu}
\authorblockA{\authorrefmark{2} Department of Electrical and Computer Engineering\\
Ohio State University\\
Columbus, OH 43202, USA \\
Email: helgamal@ece.osu.edu }}





%















\maketitle

\begin{abstract}
In this work, the role of noisy feedback in enhancing the secrecy
capacity of the wiretap channel is investigated. A model is
considered in which the feed-forward and feedback signals share
the same noisy channel. More specifically, a discrete memoryless
modulo-additive channel with a full-duplex destination node is
considered first, and it is shown that a judicious use of feedback
increases the perfect secrecy capacity to the capacity of the
source-destination channel in the absence of the wiretapper. In
the achievability scheme, the feedback signal corresponds to a
private key, known only to the destination. Then a half-duplex
system is considered, for which a novel feedback technique that
always achieves a positive perfect secrecy rate (even when the
source-wiretapper channel is less noisy than the
source-destination channel) is proposed. These results hinge on
the modulo-additive property of the channel, which is exploited by
the destination to perform encryption over the channel without
revealing its key to the source.
\end{abstract}

\section{Introduction} \label{sec:intro}

Wyner introduced the wiretap channel in~\cite{Wyner:BSTJ:75} and
established the possibility of creating an almost perfectly secure
source-destination link without relying on private (secret) keys.
In the wiretap channel, both the wiretapper and destination
observe the source encoded message through noisy channels, while
the wiretapper is assumed to have unlimited computational
resources. Wyner showed that when the source-wiretapper channel is
a degraded version of the source-destination channel, the source
can send perfectly secure messages to the destination at a
non-zero rate. The main idea is to hide the information stream in
the additional noise impairing the wiretapper by using a
stochastic encoder which maps each message to many codewords
according to an appropriate probability distribution. By doing
this, one induces maximal equivocation at the wiretapper. By
ensuring that the equivocation rate is arbitrarily close to the
message rate, one achieves perfect secrecy in the sense that the
wiretapper is now limited to learn {\em almost nothing} about the
source-destination messages from its observations.
Csis$\acute{z}$ar and K\"{o}rner generalized Wyner's approach by
considering the transmission of confidential messages over
broadcast channels~\cite{Csiszar:TIT:78}. This work characterized
the perfect secrecy capacity of the Discrete Memoryless Channel
(DMC), and showed that the perfect secrecy capacity is positive
unless the source-wiretapper channel is less noisy than the
source-destination channel (referred to as the main channel in the
sequel).

Positive secrecy capacity is not always possible to achieve in
practice. In an attempt to transmit messages securely in these
unfavorable scenarios, \cite{Maurer:TIT:93} and
\cite{Ahlswede:TIT:93} considered the wiretap channel with
noiseless feedback (The authors also considered a more general
secret sharing problem.). These works showed that one may leverage
the feedback to achieve a positive perfect secrecy rate, even when
the feed-forward perfect secrecy capacity is zero. In this model,
there exists a separate noiseless public channel, through which
the transmitter and receiver can exchange information. The
wiretapper is assumed to obtain a perfect copy of the messages
transmitted over this public channel. Upper and lower bounds were
derived for the perfect secrecy capacity with noiseless feedback
in~\cite{Maurer:TIT:93,Ahlswede:TIT:93}. In several cases, as
discussed in detail in the sequel, these bounds coincide. But, in
general, the perfect secrecy capacity with noiseless feedback
remains unknown.

Our work represents a marked departure from the public discussion
paradigm. In our model, we do not assume the existence of a
separate noiseless feedback channel. Instead, the feedback signal
from the destination, which is allowed to depend on the signal
received so far, is transmitted over the same noisy channel used
by the source. Based on the noisy feedback signal, the source can
then causally adapt its transmission scheme, hoping to increase
the perfect secrecy rate. The wiretapper receives a mixture of the
signal from the source and the feedback signal from the
destination. Quite interestingly, we show that in the
modulo-additive DMC with a full-duplex destination, the perfect
secrecy capacity with noisy feedback equals the capacity of the
main channel in the absence of the wiretapper. Furthermore, the
capacity is achieved with a simple scheme in which the source
ignores the feedback signal and the destination feeds back
randomly generated symbols from a certain finite alphabet. This
feedback signal plays the role of a private key, known only by the
destination, and encryption is performed by the modulo-additive
channel. The more challenging scenario with a half-duplex
destination, which cannot transmit and receive simultaneously, is
considered next. Here, the active transmission periods of the
destination will introduce erasures in the feed-forward
source-destination channel. In this setting, we propose a novel
feedback scheme that achieves a positive perfect secrecy rate for
any non-trivial channel distribution. The feedback signal in our
approach acts as a private {\em destination only} key which
strikes the optimal tradeoff between introducing erasures at the
destination and errors at the wiretapper. Overall, our work
proposes a novel approach for encryption where 1) the feedback
signal is used as a private key known only to the destination; and
2) the encryption is performed by exploiting the modulo-additive
property of the channel. This encryption approach is shown to be
significantly superior to the classical public discussion
paradigm.

The rest of the paper is organized as follows. In
Section~\ref{sec:model}, we introduce the system model and our
notation. Section~\ref{sec:full} describes and analyzes the
proposed feedback scheme which achieves the capacity of the full
duplex modulo-additive DMC. Taking the Binary Symmetric Channel
(BSC) as an example, we then compare the performance of the
proposed scheme with the public discussion approach. The
half-duplex scenario is studied in Section~\ref{sec:half}.
Finally, Section~\ref{sec:con} offers some concluding remarks and
outlines possible avenues for future research.

\section{The Modulo-Additive Discrete Memoryless Channel}\label{sec:model}

The modulo-additive discrete memoryless wiretap channel is
described by the following relations
\begin{eqnarray}
Y(i)=X(i)+N_1(i),\no\\
Z(i)=X(i)+N_2(i),\no
\end{eqnarray}
where, at time $i=1,\cdots,n$, $Y(i)$ is the received symbol at
the destination, $Z(i)$ is the received symbol at the wiretapper,
$X(i)$ is the channel input, $N_1(i)$ and $N_2(i)$ are the noise
samples at the destination and wiretapper, respectively. Here
$N_1(i)$ and $N_2(i)$ are allowed to be correlated, while each
process is assumed to be individually drawn from an independent
and identically distributed source. Also we have
$X(i)\in\mathcal{X}=\{0,1,\cdots,|\mathcal{X}|-1\},Y(i),N_1(i)\in\mathcal{Y}=\{0,1,\cdots,|\mathcal{Y}|-1\}$
and $Z(i),N_2(i)\in\mathcal{Z}=\{0,1,\cdots,|\mathcal{Z}|-1\}$
with finite alphabet sizes $|\mathcal{X}|,|\mathcal{Y}|$ and
$|\mathcal{Z}|$ respectively. Here `$+$' is understood to be
modulo addition with respect to the corresponding alphabet size,
i.e., $Y(i)=[X(i)+N_1(i)]\mod |\mathcal{Y}|$ and
$Z(i)=[X(i)+N_2(i)]\mod |\mathcal{Z}|$ with addition in the real
field.

In this paper, we focus on the wiretap channel with noisy
feedback. More specifically, at time $i$ the destination sends the
causal feedback signal $X_{1}(i)$ over the same noisy channel used
for feed-forward transmission, i.e., we do not assume the
existence of a separate noiseless feedback channel. The causal
feedback signal is allowed to depend on the signal received to
that point $Y^{i-1}$, i.e., $X_{1}(i)=\Psi (Y^{i-1})$, where
$\Psi$ can be any (possibly stochastic) function. In general, we
allow the destination to choose the alphabet of the feedback
signal $\mathcal{X}_1$ and the corresponding size
$|\mathcal{X}_1|$. With this {\em noisy} feedback from the
destination, the received signal at the source, wiretapper and
destination are
\begin{eqnarray}
Y_{0}(i)=X(i)+X_{1}(i)+N_0(i),\no\\
Y(i)=X(i)+X_{1}(i)+N_{1}(i),\no
\end{eqnarray}
and
$$
Z(i)=X(i)+X_{1}(i)+N_2(i),\no $$ respectively. Here
$Y_{0}(i)\in\mathcal{Y}_0=\{0,1,\cdots,|\mathcal{Y}_0|-1\}$ is the
received noisy feedback signal at the source and $N_0(i)$ is the
feedback noise, which may be correlated with $N_1(i)$ and
$N_2(i)$. We denote the alphabet size of $N_0(i)$ and $Y_0(i)$ by
$|\mathcal{Y}_0|$. Again, all `$+$' operation should be understood
to be modulo addition with corresponding alphabet size.

Now, the source wishes to send the message $W\in
\mathcal{W}=\{1,\cdots,M\}$ to the destination using an $(M,n)$
code consisting of: 1) a casual stochastic encoder $f$ at the
source that maps the message $w$ and the received noisy feedback
signal $y_0^{i-1}$ to a codeword $\mathbf{x}\in \mathcal{X}^{n}$
with
\begin{eqnarray}
x(i)=f(i,w,y_0^{i-1}),\no
\end{eqnarray}
2) a stochastic feedback encoder $\Psi$ at the destination that
maps the received signal into $X_1(i)$ with
$x_1(i)=\Psi(y^{i-1})$, and 3) a decoding function at the
destination $d$: $\mathcal{Y}^{n}\rightarrow \mathcal{W}$. The
average error probability of the $(M,n)$ code is
\begin{eqnarray}
P_{e}^{n}=\sum\limits_{w\in\mathcal{W}}\frac{1}{M}\text{Pr}\{d(\mathbf{y})\neq
w|w\text{ was sent}\}.\no
\end{eqnarray}
The equivocation rate at the wiretapper is defined as
\begin{eqnarray}
R_{e}=\frac{1}{n}H(W|\mathbf{Z}).\no
\end{eqnarray}

We are interested in achievable perfectly secure transmission
rates defined as follows.

\begin{define}
A secrecy rate $R^{f}$ is said to be achievable over the wiretap
channel with noisy feedback if for any $\epsilon>0$, there exists
an integer $n(\epsilon)$ and a sequence of codes $(M,n)$ such that
for all $n\geq n(\epsilon)$, we have
\begin{eqnarray}
R^{f}&=&\frac{1}{n}\log_2M,\label{eq:rate}\\
P_{e}^{n}&\leq& \epsilon,\label{eq:error}
\end{eqnarray}
and
\begin{eqnarray}
 \frac{1}{n}H(W|\mathbf{Z})&\geq&
R^{f}-\epsilon.\label{eq:eqvo}
\end{eqnarray}
\end{define}

\begin{define}
The secrecy capacity with noisy feedback $C_s^{f}$ is the maximal
rate at which messages can be sent to the destination with perfect
secrecy; i.e.
\begin{eqnarray}
C_s^{f}=\sup\limits_{f,d,\Psi}\{R^{f}:R^{f}\;\; \text{is
achievable}\}.\no
\end{eqnarray}
\end{define}

Note that in our model, the wiretapper is assumed to have
unlimited computational resources and to know the coding scheme of
the source and the feedback function $\Psi$ used by the
destination. We believe that our feedback model captures realistic
scenarios in which the terminals exchange information over noisy
channels.

\section{The Wiretap Channel with Full-Duplex Feedback}\label{sec:full}

\subsection{Known Results}

The secrecy capacity of the wiretap DMC without feedback $C_s$ was
characterized in~\cite{Csiszar:TIT:78}. Specializing to our
modulo-additive channel, one obtains
\begin{eqnarray}\label{eq:seccap}
C_{s}=\max\limits_{V\rightarrow X\rightarrow YZ}
[I(V;Y)-I(V;Z)]^+.
\end{eqnarray}

The wiretap DMC with public discussion was introduced and analyzed
in~\cite{Maurer:TIT:93,Ahlswede:TIT:93}. More specifically, these
papers considered a more general model in which all the nodes
observe correlated variables ( The wiretap channel model is a
particular mechanism for the nodes to observe the correlated
variables, and corresponds to the ``channel type model'' studied
in~\cite{Ahlswede:TIT:93}.), and there exists an extra noiseless
public channel with infinite capacity, through which both the
source and the destination can send information. Combining the
correlated variables and the publicly discussed messages, the
source and the destination generate a key about which the wiretap
has negligible information. Please refer to~\cite{Ahlswede:TIT:93}
for rigorous definitions of these notions. Since the public
discussion channel is noiseless, the wiretapper is assumed to
observe a noiseless version of the information transmitted over
it. It is worth noting that some of the schemes proposed
in~\cite{Maurer:TIT:93,Ahlswede:TIT:93} manage only to generate an
identical secret key at both the source and destination. The
source may then need to encrypt its message using the one-time pad
scheme which reduces the effective source-destination information
rate. Thus, the \emph{effective} secrecy rate that could be used
to transmit information from the source to the destination may be
less than the results reported
in~\cite{Maurer:TIT:93,Ahlswede:TIT:93}. Nevertheless, we use
these results for comparison purposes (which is thus generous to
the public discussion paradigm). The following theorem gives upper
and lower bounds on the secret key capacity of the public
discussion paradigm $C_s^{p}$.

\begin{thm}[\cite{Maurer:TIT:93,Ahlswede:TIT:93}] The secret key capacity of the
public discussion approach satisfies the following conditions:
\begin{eqnarray}
\max
\{\max\limits_{P_X}[I(X;Y)-I(X;Z)],\max\limits_{P_X}[I(X;Y)-I(Y;Z)]\}\no\\\leq
C_s^{p}\leq \min \{\max\limits_{P_X}I(X;Y),
\max\limits_{P_X}I(X;Y|Z)\}.\no
\end{eqnarray}
\end{thm}
\begin{proof}
Please refer to~\cite{Maurer:TIT:93,Ahlswede:TIT:93}.
\end{proof}

These bounds are known to be tight in the following
cases~\cite{Maurer:TIT:93,Ahlswede:TIT:93}.
\begin{enumerate}

\item $P_{YZ|X}=P_{Y|X}P_{Z|X}$, i.e., the main channel and the
source-wiretapper channel are independent; in this case
\begin{eqnarray}C_s^{p}=\max\limits_{P_X}\{I(X;Y)-I(Y;Z)\}.\label{eq:cwithfin}\end{eqnarray}

\item $P_{XZ|Y}=P_{X|Y}P_{Z|Y}$, i.e., $X\rightarrow Y\rightarrow
Z$ forms a Markov chain, and hence the source-wiretapper channel
is a degraded version of the main channel. In this case
\begin{eqnarray}
C_{s}^{p}=\max\limits_{P_X}\{ I(X;Y)-I(X;Z)\}\label{eq:cwithfdeg}.
\end{eqnarray}

This is also the secrecy capacity of the degraded wiretap channel
without feedback. Hence public discussion does not increase the
secrecy capacity for the degraded wiretap channel.

\item $P_{XY|Z}=P_{X|Z}P_{Y|Z}$, i.e., $X\rightarrow Z\rightarrow
Y$, so that the main channel is a degraded version of the wiretap
channel. In this case
\begin{eqnarray}\label{eq:maindegrad}
C_s^{p}=0.
\end{eqnarray}

Again, public discussion does not help in this scenario.
\end{enumerate}

\subsection{The Main Result}

The following theorem characterizes the secrecy capacity of the
wiretap channel with noisy feedback. Moreover, achievability is
established through a novel encryption scheme that exploits the
modulo-additive structure of the channel and uses a private key
known only to the destination.

\begin{thm}\label{thm:dmc}
The secrecy capacity of the discrete memoryless modulo-additive
wiretap channel with noisy feedback is
\begin{eqnarray}C_s^{f}=C,\no\end{eqnarray}
where $C$ is the capacity of the main channel in the absence of
the wiretapper.
\end{thm}
\begin{proof}(Sketch) For the converse part, we first upper bound the secrecy capacity
with feedback by the channel capacity with feedback in the absence
of the wiretapper, by ignoring the equivocation
condition~\eqref{eq:eqvo}. Since feedback does not increase the
channel capacity of the DMC, the channel capacity with feedback in
the absence of the wiretapper is $C$, which completes the converse
part.

For the achievability part, we use the following novel scheme. The
source ignores the feedback signal and uses a channel coding
scheme for the ordinary channel without a wiretapper. The
destination sets $\mathcal{X}_1=\mathcal{Z}$, and at any time $i$
sets $X_{1}(i)=a$, where $a$ is chosen randomly and equiprobably
from $\{0,\cdots,|\mathcal{Z}|-1\}$. Hence $\mathbf{X}_1$ is
uniformly distributed over $\mathcal{Z}^n$. After receiving
$\mathbf{Y}$, the destination first sets
$\hat{\mathbf{Y}}=\mathbf{Y}-\mathbf{X}_1$, and then obtains an
estimate of the codeword by finding an $\hat{\mathbf{X}}$ that is
jointly typical with $\hat{\mathbf{Y}}$. It can be shown that the
error probability at the destination can be made arbitrarily
small. The wiretapper receives
$\mathbf{Z}=\mathbf{X}+\mathbf{X}_1+\mathbf{N}_2$, which can be
shown to be uniformly distributed over $\mathcal{Z}^n$ for any
realizations of $\mathbf{X}$ and $\mathbf{N}_2$. Hence
$I(X;\mathbf{Z})=0$, which further implies $I(W;\mathbf{Z})=0$ by
the data processing inequality.

Please refer to~\cite{Lai:TIT:07} for further details.
\end{proof}

Our scheme achieves $I(W;\mathbf{Z})=0$. This implies perfect
secrecy in the strong sense of Shannon~\cite{Shannon:BSTJ:49} as
opposed to Wyner's notion of perfect secrecy~\cite{Wyner:BSTJ:75},
which has been pointed out to be insufficient for certain
cryptographic applications~\cite{Maurer:LNCS:00}. The enabling
observation behind our achievability scheme is that, by
judiciously exploiting the modulo-additive structure of the
channel, one can render the channel output at the wiretapper
independent of the codeword transmitted by the source. Here, the
feedback signal $\mathbf{X}_1$ serves as a private key and the
encryption operation is carried out by the channel. Instead of
requiring both the source and destination to know a common
encryption key, we show that only the destination needs to know
the encryption key, hence eliminating the burden of secret key
distribution. Remarkably, the secrecy capacity with {\em noisy}
feedback is shown to be larger than the secret key capacity of
public discussion schemes. This point will be further illustrated
by the binary symmetric channel example discussed next. This
presents a marked departure from the conventional wisdom, inspired
by the data processing inequality, which suggests the superiority
of noiseless feedback. This result is due to the fact that the
noiseless feedback signal is also available to the wiretapper,
while in the proposed noisy feedback scheme neither the source nor
the wiretapper knows the feedback signal perfectly. In fact, the
source in our scheme ignores the feedback signal, which is used
primarily to {\em confuse} the wiretapper. Our result shows that
complicated feedback functions $\Psi$ are not needed to achieve
optimal performance in this setting (i.e., a random number
generator suffices). Also, the alphabet size of the feedback
signal can be set equal to the alphabet size of the wiretapper
channel and the coding scheme used by the source is the same as
the one used in the absence of the wiretapper.


\subsection{The Binary Symmetric Channel Example}
\begin{figure}[thb]
\centering
\includegraphics[width=0.2\textwidth]{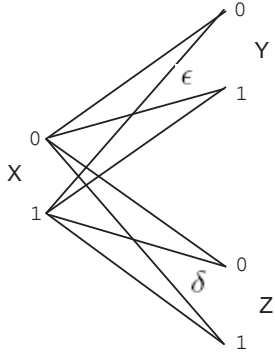}
\caption{The Binary Symmetric Wiretap Channel.} \label{fig:bsc}
\end{figure}
To illustrate the idea of encryption over the channel, we consider
in some details the wiretap BSC shown in Figure~\ref{fig:bsc},
where $\mathcal{X}=\mathcal{Y}=\mathcal{Z}=\{0,1\}$,
$\text{Pr}\{N_1(i)=1\}=\epsilon$ and
$\text{Pr}\{N_2(i)=1\}=\delta$. The secrecy capacity of this
channel without feedback is known to be~\cite{Maurer:TIT:93}
$$C_s=[H(\delta)-H(\epsilon)]^+,$$ with $H(x)=-x\log x-(1-x)\log
(1-x)$. We differentiate among the following special cases.

\begin{enumerate}

\item $\epsilon=\delta=0$.

In this case, both the main channel and wiretap channel are
noiseless, hence $C_s=0.$ Also we have $C_s^p=0,$ since the
wiretapper sees exactly the same as what the destination sees.
Specializing our scheme to this BSC channel, at time $i$, the
destination randomly chooses $X_1(i)=1$ with probability 1/2 and
sends $X_1(i)$ over the channel. This creates a virtual BSC at the
wiretapper with $\delta^{'}=1/2$. On the other hand, since the
destination knows the value of $X_1(i)$, it can cancel it by
adding $X_1(i)$ to the received signal. This converts the original
channel to an equivalent BSC with $\epsilon^{'}=0$. Hence, through
our noisy feedback approach, we obtain an equivalent wiretap BSC
with parameters $\epsilon^{'}=0$ and $\delta^{'}=1/2$ resulting in
$C_s^{f}=H(\delta^{'})-H(\epsilon^{'})=1.$

\item $0<\delta<\epsilon<1/2$, $N_1(i)$ and $N_2(i)$ are
independent.

Since $\delta<\epsilon$, we have $C_s=0.$ Also, $N_1(i)$ and
$N_2(i)$ are independent, so $P_{YZ|X}=P_{Y|X}P_{Z|X}$. Then
from~\eqref{eq:cwithfin}, one can easily obtain
that~\cite{Maurer:TIT:93}
$C_s^{p}=H(\epsilon+\delta-2\epsilon\delta)-H(\epsilon).$

Our feedback scheme, on the other hand, achieves
$C_s^{f}=1-H(\epsilon).$ Since
$H(\epsilon+\delta-2\epsilon\delta)\leq 1$, we have $C_s^{f}\geq
C_s^{p}$ with equality if and only if
$\epsilon+\delta-2\epsilon\delta=1/2$.

\item $0<\delta<\epsilon<1/2$ and $N_1(i)=N_2(i)+N^{'}(i)$, where
$\text{Pr}\{N^{'}(i)=1\}=(\epsilon-\delta)/(1-2\delta)$.

The main channel is a degraded version of the source-wiretapper
channel, $X\rightarrow Z\rightarrow Y$, as shown in
Figure~\ref{fig:maindegrad}.

\begin{figure}[thb]
\centering
\includegraphics[width=0.4\textwidth]{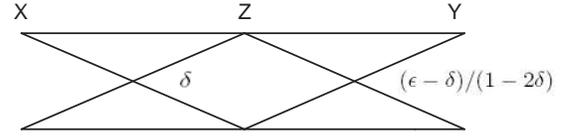}
\caption{The BSC Wiretap Channel with a Degraded Main Channel.}
\label{fig:maindegrad}
\end{figure}

Hence, from~\eqref{eq:maindegrad}, we have
$C_s=C_s^{p}=0,$
while $C_s^{f}=1-H(\epsilon).$

 \item $0<\epsilon<\delta<1/2$, and
$N_2(i)=N_1(i)+N^{'}(i)$, where
$\text{Pr}\{N^{'}(i)=1\}=(\delta-\epsilon)/(1-2\epsilon)$.

\begin{figure}[thb]
\centering
\includegraphics[width=0.4\textwidth]{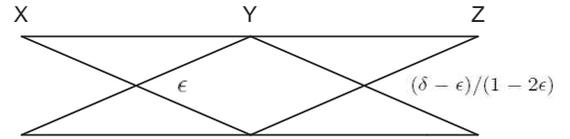}
\caption{The BSC wiretap Channel with a Degraded Source-Wiretapper
Channel.} \label{fig:wiretapdegrad}
\end{figure}

In this case, the source-wiretapper channel is a degraded version
of the main channel as shown in Figure~\ref{fig:wiretapdegrad};
$X\rightarrow Y\rightarrow Z$, so from~\eqref{eq:cwithfdeg}
$C_s=C_s^{p}=H(\delta)-H(\epsilon).$
But $C_s^{f}=1-H(\epsilon)\geq C_s^{p}$ with equality if and only
if $\delta=1/2$.

\item $N_1(i)$ and $N_2(i)$ are correlated and the channel is not
degraded.

In this case
$C_s=[H(\delta)-H(\epsilon)]^+.$
The value of $C_s^p$ is unknown in this case but can be bounded by
\begin{eqnarray}
C_s=[H(\delta)-H(\epsilon)]^+\leq C^p_s\leq
1-H(\epsilon)=C_s^{f}.\no
\end{eqnarray}

\end{enumerate}
In summary, the secrecy capacity with noisy feedback is always
larger than or equal to that of the public discussion paradigm
when the underlying wiretap channel is a BSC. More strongly, the
gain offered by the noisy feedback approach, over the public
discussion paradigm, is rather significant in many relevant
special cases.

\section{Even Half-duplex Feedback is Sufficient}\label{sec:half}

It is reasonable to argue against the practicality of the full
duplex assumption adopted in the previous section. For example, in
the wireless setting, nodes may not be able to transmit and
receive with the same {\em degree of freedom} due to the large
difference between the power levels of the transmit and receive
chains. This motivates extending our results to the half duplex
wiretap channel in which the terminals can either transmit or
receive but never both at the same time. Under this situation, if
the destination wishes to feed back at time $i$, it loses the
opportunity to receive the $i^{th}$ symbol transmitted by the
source, which effectively results in an erasure (assuming that the
source is unaware of the destination's decision). The proper
feedback strategy must, therefore, strike a balance between
confusing the wiretapper and degrading the source-destination
link. In order to simplify the following presentation, we first
focus on the wiretap BSC. The extension to arbitrary
modulo-additive channels is briefly outlined afterwards.

In the full-duplex case, at any time $i$, the optimal scheme is to
let the destination send $X_1(i)$, which equals $0$ or $1$
equiprobably. But in the half-duplex case, if the destination
always keeps sending, it does not have a chance to receive
information from the source, and hence, the achievable secrecy
rate is zero. This problem, however, can be solved by observing
that if at time $i$, $X_1(i)=0$, the signal the wiretapper
receives, i.e., $Z(i)=X_1(i)+N_2(i),$ is the same as in the case
in which the destination does not transmit. The only crucial
difference in this case is that the wiretapper does not know
whether the feedback has taken place or not, since $X_1(i)$ can be
randomly generated at the destination and kept private.

The previous discussion inspires the following feedback scheme for
the half-duplex channel. The destination first fixes a value $t\in
[0,1]$ which is revealed to both the source and wiretapper. At
time $i$, the destination randomly generates $X_1(i)=1$ with
probability $t$ and $X_1(i)=0$ with probability $1-t$. If
$X_1(i)=1$, the destination sends $X_1(i)$ over the channel, which
causes an erasure at the destination and a potential error at the
wiretapper. On the other hand, when $X_1(i)=0$, the destination
does not send a feedback signal and spends the time on receiving
from the channel. The key to this scheme is that although the
source and wiretapper know $t$, neither is aware of the exact
timing of the event $\{X_1(i)=1\}$. The source ignores the
feedback and keeps sending information. The following result
characterizes the achievable secrecy rate with the proposed
feedback scheme.

\begin{thm}
For a BSC with half-duplex nodes and parameters $\epsilon$ and
$\delta$, the scheme proposed above achieves
\begin{eqnarray}
R_s^{f}&&\hspace{-7mm}=\max\limits_{\mu,t}\left[(1-t)\big[H(\epsilon+\mu-2\mu\epsilon)-H(\epsilon)\big]-\right.\no\\
&&\hspace{20mm}\left.\big[H(\hat{\delta}+\mu-2\mu\hat{\delta})-H(\hat{\delta})\big]\right]^+,\no
\end{eqnarray}
with $\hat{\delta}=\delta+t-2\delta t$.
\end{thm}

In general, one can obtain the optimal values of $\mu$ and $t$ by
setting the partial derivatives of $R^{f}$, with respect to $\mu$
and $t$ to 0, and solving the corresponding equations.
Unfortunately, except for some special cases, we do not have a
closed form solution for these equations at the moment.
Interestingly, using the {\em not necessarily optimal} choice of
$\mu=t=1/2$, we obtain $R^{f}=[1-H(\epsilon)]/2$ implying that we
can achieve a nonzero secrecy rate as long as $\epsilon\neq 1/2$
irrespective of the wiretapper channel conditions. Hence, even for
half-duplex nodes, noisy feedback from the destination allows for
transmitting information securely for {\em almost} any wiretap
BSC. Finally, we compare the performance of different schemes in
some special cases of the wiretap BSC.
\begin{enumerate}

\item $\epsilon=\delta=0$.

As mentioned above, here we have $C_s=C_s^{p}=0$. It is easy to
verify that the optimal choice of $\mu$ and $t$ are $1/2$, and we
thus have $R_s^{f}=1/2$.

\item $0<\delta<\epsilon<1/2$ and $N_1(i)=N_2(i)+N^{'}(i)$, where
$\text{Pr}\{N^{'}(i)=1\}=(\epsilon-\delta)/(1-2\delta)$.

The main channel is a degraded version of the wiretap channel, so
$ C_s=C_s^{p}=0.$ But by setting $\mu=t=1/2$ in our half-duplex
noisy feedback scheme, we obtain $R_s^{f}=[ 1-H(\epsilon)]/2.$
\end{enumerate}

The extension to the general discrete modulo-additive channel is
natural. The destination can set $\mathcal{X}_1=\mathcal{Z}$, and
generates $X_1(i)$ with a certain distribution $P_{X_1}$. At time
$i$, if the randomly generated $X_1(i)\neq 0$, the destination
sends a feedback signal, incurring an erasure to itself. On the
other hand, if $X_1(i)=0$, it does not send the feedback signal
and spends the time listening to the source. The achievable
performance could be calculated based on the equivalent channels
as done in the BSC. This scheme guarantees a positive secrecy
capacity as seen in the case where $P_{X_1}$ is chosen to be
uniformly distributed over $\mathcal{Z}$. This is because a
uniform distribution over $\mathcal{Z}$ renders the output at the
wiretapper independent from the source input, i.e., $I(W;Z)=0$,
while the destination can still spend $1/|\mathcal{Z}|$ part of
the time listening to the source. Finding the optimal distribution
$P_{X_1}$, however, is tedious.

\section{Conclusion}\label{sec:con}
In this paper, we have obtained the secrecy capacity (or
achievable rate) for several instantiations of the wiretap channel
with noisy feedback. More specifically, with a full duplex
destination, it has been shown that the secrecy capacity of
modulo-additive channels is equal to the capacity of the
source-destination channel in the absence of the wiretapper.
Furthermore, the secrecy capacity is achieved with a simple scheme
in which the destination randomly chooses its feedback signal from
a certain finite alphabet. Interestingly, with a slightly modified
feedback scheme, we are able to achieve a positive secrecy rate
for the half duplex channel. We have shown that this paradigm
significantly outperforms the public discussion approach for
sharing private keys between the source and destination. Our
results motivate several interesting directions for future
research. For example, characterizing the secrecy capacity of
arbitrary DMCs (and the AWGN channel) with feedback remains an
open problem. From an algorithmic perspective, it is also
important to understand how to exploit different channel
structures (in addition to the modulo-additive one) for encryption
purposes. Finally, extending our work to multi-user channels
(e.g., the relay-eavesdropper channel~\cite{Lai:TIT:061}) is of
considerable interest.


\end{document}